\documentclass[prl,aps,twocolumn,showpacs,amsmath,amssymb]{revtex4}

\usepackage[dvips]{graphicx}
\usepackage{dcolumn}% Align table columns on decimal point
\usepackage{bm}% bold math

\begin{document}

\preprint{}

\title{Scale-free trees: the skeletons of complex networks}

\author{Dong-Hee Kim$^{1}$, Jae Dong Noh$^{2}$, and Hawoong Jeong$^{1}$}
\affiliation{$^{1}$Department of Physics,
Korea Advanced Institute of Science and Technology, 
Daejeon 305-701, Korea}
\affiliation{$^{2}$Department of Physics,
Chungnam National University, Daejeon 305-764, Korea}

\date{\today}

\begin{abstract}
We investigate the properties of the spanning trees of 
various real-world and model networks. The spanning tree 
representing the communication kernel of the original network 
is determined by maximizing total weight of edges, whose weights are 
given by the edge betweenness centralities. 
We find that a scale-free tree and shortcuts organize a complex network.
The spanning tree shows robust betweenness centrality 
distribution that was observed in scale-free tree models.
It turns out that the shortcut distribution characterizes the properties of 
original network, such as the clustering coefficient and 
the classification of networks by the betweenness centrality distribution.

\end{abstract}

\pacs{89.75.Hc, 89.75.Da, 89.75.Fb, 05.10.-a}

%\keywords{}

\maketitle

Complex network theories have attracted much attention in 
last few years with the advance in the understanding of the highly 
interconnected nature of various social, biological and 
communication systems \cite{Albert1,Dorogovtsev1}. 
The inhomogeneity of network structures 
is conveniently characterized by the degree distribution $P_d(k)$ 
, the probability for a vertex to have $k$ edges toward other vertices. 
The emergence of {\it scale-free} distribution 
$P_d(k) \sim k^{-\gamma}$ has been reported in many real-world networks, 
such as the coauthorship networks in social systems \cite{Newman1}, the 
metabolic networks and the protein interaction networks in biological systems
\cite{Jeong1,Jeong2}, and Internet and World Wide Web 
in technological systems \cite{Faloutsos1,Albert2}.

It is important to study the dynamics on networks as well as to study 
the structural properties of networks since its application to 
the real-world. However, the dynamical phenomena on networks 
such as traffic and information flow are very difficult to predict 
from local information due to rich microstructures and 
corresponding complex dynamics.
Thus, to understand the dynamical phenomena on networks, one must know 
the global properties of networks as well as the local properties
such as degree distribution.  
It is the reason why the dynamics on complex networks has not been studied 
systematically so far.

Due to their inhomogeneous structure,
traffic or information flow on complex networks would be also very 
inhomogeneous. As a simplified quantity to measure the traffic 
of networks, it is natural to use the betweenness centrality (BC) 
\cite{Freeman1,Girvan1,Newman2}.
The BC of $\mathbf{G}$, either a vertex or an edge, is defined as 
\begin{equation}
b(\mathbf{G}) = \sum_{i \neq j} b(i,j;\mathbf{G})  
= \sum_{i \neq j} \frac{c(i,j;\mathbf{G})}{c(i,j)},
\end{equation}
where $c(i,j;\mathbf{G})$ denotes 
the number of shortest paths from a vertex $i$ to $j$ 
through $\mathbf{G}$, and $c(i,j)$ is 
the total number of shortest paths from $i$ to $j$.
In terms of the packet in the Internet, assuming every vertex sends 
a unit packet to each of other vertices, BC is the average amount of 
packets passing though a vertex or an edge.

In scale-free networks, the distribution of the vertex BC is known 
to follow the power-law with the exponents of 
either $2.2$ or $2.0$ \cite{Goh3}. Though the edge BC 
distribution does not follow power-law exactly, 
the distribution of the edge BC is also 
very inhomogeneous in scale-free networks \cite{dhkim1}. 
This indicates that there exist extremely essential edges 
having large edge BCs which are used for communication very frequently. 
Thus, one can imagine a sub-network constructed only by the 
essential edges with global connectivity retained. 
We regard this network as a communication kernel, 
which handles most of the traffic on a network.

For simplicity, we define the communication kernel of a network as the 
spanning tree with a set of edges maximizing the 
summation of their edge BCs on the original networks.
The constructing procedure is very similar to the 
minimum spanning tree algorithm \cite{Kruskal1}. 
We repeatedly select an edge according to the priority of the edge BC,
and add the edge to the tree if it does not make any loop 
until the tree includes all vertices \cite{comment1}.
Note that the residual edges can be regarded as the shortcuts 
since they shorten paths on the spanning tree. This concept 
of the spanning tree and shortcuts corresponds 
to that of 1-D regular lattice and shortcuts, respectively, in 
the small-world networks \cite{Watts1}.

In this paper, we investigate the structural and 
dynamical properties of the spanning tree of 
complex networks and the role of shortcuts in the networks. 
For various scale-free real-world and model networks,
we find that the spanning trees show the scale-free behavior.
We also find that the vertex and edge BC distributions 
follow the power-law with the robust exponent $\eta = 2.0$,
regardless of the exponent value $\eta = 2.0$ or $2.2$ of 
original networks.
On the other hand, it turns out that the shortcut length distribution
shows either Gaussian-like or monotonically decaying behavior
depending on the BC distribution exponent $\eta$ of original networks.

\begin{table*}
\begin{ruledtabular}
\begin{tabular}{lcccccccccccc}
Network & N & $\langle k \rangle$ & $f$ & $f_0$ & $\gamma$ & $\gamma_s$ &
$\eta$ & $\eta_s$ & $r$ & $r_s$ & $r_p$ & ref. \\
\hline
NEURO & 190382 & 12.5 & 0.46 & 0.16 & 2.1(1) & 2.4(1) 
& 2.2(1) & 2.0(1) & 0.601 & -0.138 & 0.538 & \cite{Barabasi2}\\
arxiv.org & 44336 & 10.79 & 0.54 & 0.19 & - & 2.1(1) 
& - & 2.0(1) & 0.352 & -0.119 & 0.497 & \cite{Newman2}\\
arxiv.org/cond-mat & 13860 & 6.43 & 0.61 & 0.31 & - & 2.7(1) 
& - & 2.0(1) & 0.157 & -0.187 & 0.714&\cite{Newman2}\\
PIN & 4926 & 6.55 & 0.54 & 0.3 & - & 2.3(1) 
& 2.3(1) & 2.0(1) & -0.139 & -0.161 & 0.814 & \cite{Goh1} \\
BA model & $2 \times 10^5$ & 4 & 0.71 & 0.5 & 3.0(1) & 2.7(1) 
& 2.2(1) & 2.0(1) & $\sim 0$ & $\sim 0$ & 0.973 &\cite{Barabasi1} \\ 
Holme-Kim model & $10^4$ & 6 & 0.58-0.71 & 0.33 & 3.0(1) & 2.4(1) 
& 2.2(1) & 2.0(1) & -0.033 & -0.117 & 0.947 & \cite{Holme1} \\
Static model & $\sim 10^4$ & $\sim 4$ & 0.65 & $\sim 0.5$ & 2.6-3.0 & 2.4-2.8
& 2.2(1) & 2.0(1) & -0.022 & -0.067 & 0.938 & \cite{Goh4} \\
Fitness model & $2 \times 10^5$ & 4 & 0.73 & 0.5 & 2.25 & 2.2(1) 
& 2.2(1) & 2.0(1) & $\sim 0$ & $\sim 0$ & 0.994 & \cite{Bianconi1}\\
Internet AS & 10514 & 4.08 & 0.65 & 0.5 & 2.1(1) & 2.1(1) 
& 2.0(1) & 2.0(1) & -0.185 & -0.183 & 0.929 & \cite{Meyer1} \\ 
Adaptation model & $ \sim 10^5 $ & 11.9 & 0.503 & 0.17 & 2.1(1) & 2.1(1) 
& 2.0(1) & 2.0(1) & -0.219 & -0.215 & 0.749 & \cite{Goh2} \\
\end{tabular}
\end{ruledtabular}
\caption{\label{table1}The scaling exponents and correlation coefficients
of the spanning trees and original networks for various real-world networks 
and models. Tabulated for each network is the system size $N$, 
the mean degree $\langle k \rangle$, the ratio of edge BC 
summation over the edges selected for the spanning tree to 
total edge BC $f$, the ratio of the number of edges in spanning trees 
and original networks $f_0$, the degree exponent $\gamma$, 
the BC exponent $\eta$, the assortativity $r$, 
and the degree correlation coefficient $r_p$ between the original network 
and the spanning tree. The $s$ subscripts indicate quantities for 
the spanning trees.
Here we consider only the largest cluster of networks 
when network has several disjoint parts.}
\end{table*}

Firstly, we confirm the spanning tree to be a communication kernel 
by estimating the relative importance of selected edges in 
the obtained spanning tree and those from the random selection. 
If we select the edges randomly, the fraction $f$ of 
the edge BC summation over the selected edges and 
that over total edge would be approximately 
$f_0$, the ratio of the number 
of edges in the tree and that of network.
However, it turns out that the real set of selected edges from 
the spanning tree possesses over $50 \%$ of 
the total edge BC of the network (see Table~\ref{table1}.), 
therefore $f \gg f_0$. 
For instance, the coauthorship network shows 
that $f$ is nearly three times larger than $f_0$ even though 
the number of edges in the spanning tree is only $16 \%$ of 
that in the original network. 
Thus we can call this spanning tree the communication kernel.

\begin{figure}
\includegraphics[width=0.40\textwidth]{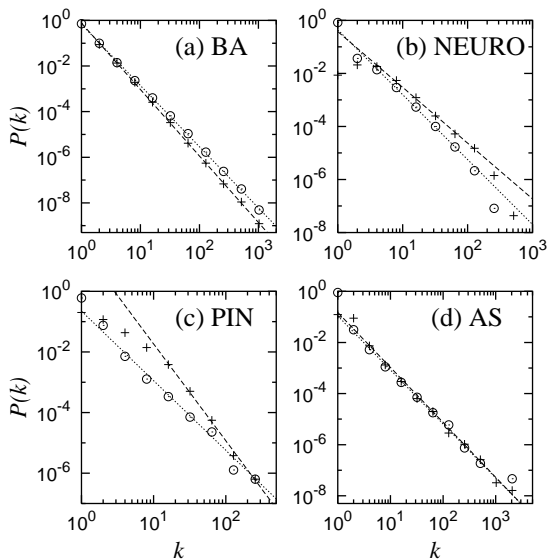}
\caption{\label{fig1}Degree distributions of 
the spanning trees ($\bigcirc$) and their original networks ($+$), 
(a) BA model with $m=2$, 
(b) coauthorship network, NEURO, (c) 
PIN, and (d) Internet AS. The data 
points are shifted vertically to enhance the visibility.}
\end{figure}

To find out more about this kernel, we measure the degree distribution 
of the spanning trees. It turns out that the degree distribution 
always follows the power-law \cite{comment2}, 
which is tested for various networks including 
the Barab\'asi-Albert (BA) model \cite{Barabasi1}, 
coauthorship network in neuroscience (NEURO) \cite{Barabasi2}, 
protein interaction networks of yeast (PIN) \cite{Goh1}, 
Internet at the autonomous systems (AS) level \cite{Meyer1},
and so on (see Fig.~\ref{fig1} and Table~\ref{table1}).
However, the details of the degree distribution depend on
each of the networks. 
The exponents of the power-law degree distributions 
of the spanning trees do not always agree with those of the original networks
(see Table~\ref{table1}). This indicates that 
the spanning trees are far from the random sampling of edges. 

\begin{figure}
\includegraphics[width=0.40\textwidth]{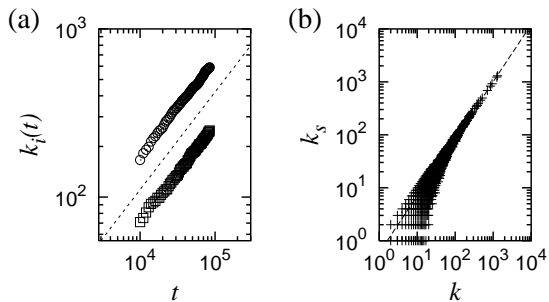}
\caption{\label{fig2}(a) Time evolution for the degree 
of two vertices added to system at $t=5$ ($\bigcirc$) 
and $t=55$ ($\square$), where the dashed
line is a linear fit with slope 0.58. 
(b) Scattered plot for degree of the original 
network ($k$) and the spanning tree ($k_s$). The dashed line has 
the slope of $1.08$.}
\end{figure}

To confirm the scale-free behavior of the spanning tree,
we investigate the time evolution of the degree in a growing network.
Assuming that the fixed number of new 
vertices are introduced at each time step in growing networks, 
it is well known that the degree following the power-law 
$k_i(t) \sim t^{\beta}$ leads to the scale-free 
degree distribution $P_d(k) \simeq k^{-\gamma}$ \cite{Barabasi3}, 
where $k_i(t)$ is the degree of the vertex $i$ at time $t$ and 
$\gamma = 1/\beta +1$. This argument can be naturally applied for 
the spanning tree of the BA model since it grows constantly. 
At each time step of the growth 
in the BA model, we obtain the spanning tree and 
measure the degree of every vertex.
In Fig.~\ref{fig2}(a), we show the time evolution of the 
degrees of several vertices. The degrees evolve with 
$\beta = 0.58$ that leads $\gamma_s=2.7$ of 
the spanning tree, which agrees with our measurement 
from the actual degree distribution.

The high correlation between the degrees from spanning trees 
and the original networks also guarantees the preserved 
scale-free behavior of the spanning trees.
The correlation coefficient between the degree of the original 
network $k$ and the degree of its spanning trees $k_s$ is defined as 
the Pearson's correlation coefficient between $k$ and $k_s$, 
$ r_p = \frac{\langle k k_s \rangle - \langle k \rangle \langle k_s \rangle}
{\sqrt{( \langle k^2 \rangle - \langle k \rangle^2 )
(\langle k_s^2 \rangle - \langle k_s \rangle^2 )}} $.
Most networks exhibit strong degree correlation between the
spanning tree and its original network (see Table~\ref{table1}). 
We find that the degrees of networks ($k$) and their spanning trees ($k_s$) 
roughly follow $k_s \sim k^{\alpha}$ that leads the 
degree distribution of the spanning trees $P_d(k_s) \sim k_s^{-\gamma_s}$ 
with $\gamma_s =  (\gamma + \alpha -1)/\alpha$. 
In Fig.~\ref{fig2}(b), we show that the BA model has $\alpha = 1.08$,
which leads $\gamma_s = 2.75$ in good agreement 
with the result obtained by direct measurement.

The assortativity is another interesting feature of the spanning trees.
The assortativity $r$ \cite{Newman3}, that measures the degree correlation 
of vertices directly connected by an edge, is defined by 
$r = \frac{ \langle jk \rangle - \langle j \rangle \langle k \rangle }
{ \langle k^2 \rangle - \langle k \rangle^2 }$,
where $j$ and $k$ are the remaining degrees at the end of an edge and 
the angular brackets indicate the average over all edges. 
We find that all spanning trees show dissortative or neutral 
behavior regardless of the assortativity of original networks 
(see Table \ref{table1}). Thus, we can propose that it is 
general characteristics of the spanning trees of scale-free networks. 
We need further study to prove our conjecture.

\begin{figure}
\includegraphics[width=0.40\textwidth]{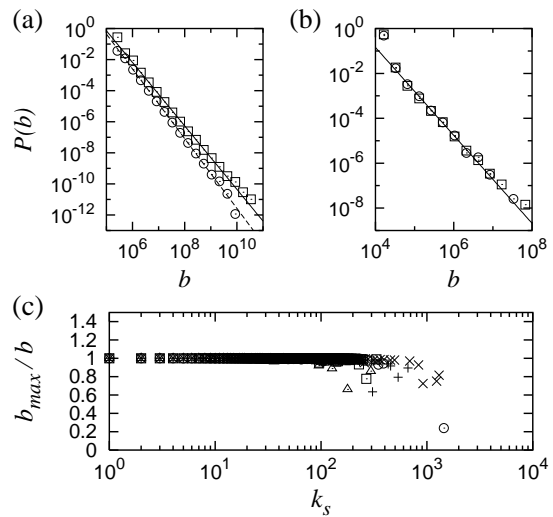}
\caption{\label{fig3}The vertex BC distribution 
of the original networks ($\bigcirc$) and the spanning tree ($\square$) 
for (a) the BA model averaged over 10 ensembles 
and (b) Internet AS. In (a), the solid and dashed line have 
the slopes of $2.0$ and $2.2$, respectively. 
The lines in (b) are linear fits with slope $2.0$. 
The data points are shifted vertically to enhance the visibility.
(c) The ratio of the largest value of edge BC ($b_{max}$)
to vertex BC ($b$) of a vertex with degree $k_s$ for 
the BA tree ($+$) and the spanning trees of the BA model ($\times$), 
NEURO ($\square$), Internet AS ($\bigcirc$), and 
PIN ($\bigtriangleup$) networks.}
\end{figure}

We find that the BC distribution of the spanning tree 
is robust regardless of its original networks.
For both of vertices and edges, the BC distribution follow
the power-law with the robust exponent $\eta_s = 2.0$ 
in all spanning trees we studied (see Fig.~\ref{fig3} and Table~\ref{table1}).
This is consistent with the numerical results for the 
known scale-free tree models \cite{Goh3}. 
The same BC distribution for vertices and edges is the general 
feature of trees. In the mean field picture, 
the largest BC of edges belonging to a vertex gives 
dominant contribution to the BC of the vertex \cite{Szabo1}. 
For our obtained spanning trees, we verify numerically 
that the largest edge BC of a vertex almost equals to the vertex BC 
for most of vertices (See Fig.~\ref{fig3}(c)). 

\begin{figure}
\includegraphics[width=0.40\textwidth]{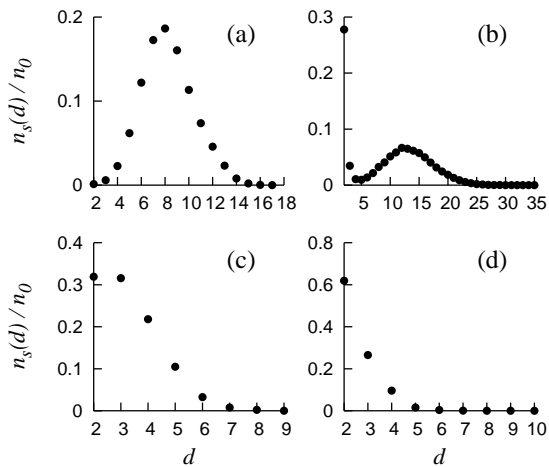}
\caption{\label{fig4} The length distribution of shortcuts for 
(a) the BA model (m=2), (b) coauthorship network of 
neuroscience, (c) Internet AS, and (d) adaptation model with $10^5$ vertices. 
$n_s(d)$ and $n_0$ are the number of shortcuts with the length $d$ and 
total number of shortcuts, respectively.}
\end{figure}

The spanning trees show the robust features, such as scale-free degree 
distribution, robust BC distribution, and dissortative or 
neutral degree correlation. Here one can ask 
what is the role of shortcuts which are not included in the 
spanning tree. To answer this question, 
we focus on the length of the shortcuts on the spanning trees.
The length of a shortcut between vertices $i$ and $j$ is defined as 
the minimum number of hops from $i$ to $j$ on the spanning tree.
The non-zero clustering coefficient of the original networks 
can now be explained by short-length shortcuts. 
Obviously, shortcuts with the length $2$ 
build triangles of vertices, hence increase the clustering coefficient.
All networks with non-vanishing clustering coefficient have the 
significant amount of the shortcuts with the length $2$ (see Fig.~\ref{fig4}).  

Interestingly, we find that there are two types in 
the shortcut length distribution (see Fig.~\ref{fig4}). 
In one distribution (Type I), most shortcuts distribute 
near a large mean value, similar to the Gaussian distribution, 
which shows that the network is the longer-loop dominant structure. 
In the other distribution (Type II), the number of 
shortcuts monotonically decreases as the length 
increases, which indicates that the network is tree-like.
Most of networks including the BA model, coauthorship networks, 
and PIN belong to the type I. On the other hand, Internet AS and 
the adaptation model are type II. 
We find that our classification exactly agrees with the grouping by 
the exponent of the BC distribution \cite{Goh3}.
The networks belonging to type I or type II show  
vertex BC distributions with the exponents of 
$2.2$ or $2.0$, respectively. 
Goh, \textit{et al.} \cite{Goh3} characterized the networks with 
the BC exponent of $2.0$ as the linear mass-distance relation, 
which shows that the shortest paths of the networks are 
similar to trees. Our result also supports the tree-like structures
of the type II networks and give an intuitive explanation
of the reason why the BC exponents of the type II networks are 
as same as those of the scale-free trees.
Because there exists mostly short length shortcuts in the type II 
networks with monotonically fast-decaying shortcut length distribution, 
the structure of the original networks are not significantly 
different from their spanning trees. Therefore, 
The BC exponents of the type II networks are unchanged at $2.0$
of their spanning trees.

In summary, we study the properties of the spanning trees with 
maximum total edge betweenness centrality, which is regarded as
the communication kernel on networks. We find that 
a complex network can be decomposed into a scale-free tree and 
additional shortcuts on it. The scale-free trees show robust 
characteristics in the betweenness centrality distribution and 
the degree correlation. The remaining shortcuts are responsible for the 
detailed characteristics of the networks such as the clustering 
property and the BC distribution. 
The distribution of the shortcut length clearly distinguishes
the network into the two types, which coincides with the classes 
determined from the BC exponents \cite{Goh3}.

The authors are grateful to Byungnam Kahng, Kwang-Il Goh, and 
Mark Newman for the fruitful discussions and supporting 
the real-world network data.
This work was supported by grant No. 
R14-2002-059-01002-0 from KOSEF-ABRL program and
by the Korean Systems Biology Research Grant 
(M10309020000-03B5002-00000) from the Ministry of Science and Technology.


\begin{thebibliography}{}
\bibitem{comment1} We study only undirected networks for simplicity of 
algorithm. However, our analysis can be generally applied to directed
networks.
\bibitem{comment2} We also observed the scale-free behavior in 
the spanning trees constructed using other methods, 
such as percolation approach and shortest path finding 
algorithm, which will be published elsewhere.
\end{thebibliography}
\end{document}